\documentclass{elsart}
\newcommand{\be}{\begin{equation}}
\newcommand{\ee}{\end{equation}}
\newcommand{\lb}[1]{\label{#1}}

\begin{document}
\runauthor{Moura Jr.\ and Ribeiro}
\begin{frontmatter}
\title{Zipf Law for Brazilian Cities}
\author[NJMJr]{Newton J.\ Moura Jr}
\author[MBR]{and Marcelo B.\ Ribeiro}

\address[NJMJr]{IBGE -- Brazilian Institute for Geography and Statistics,
              Geosciences Directorate, Geodesics Department,
              Av.\ Brasil 15671, Rio de Janeiro, RJ 21241-051, Brazil;
	      e-mail:~newtonjunior@ibge.gov.br}
\address[MBR]{Physics Institute, University of Brazil -- UFRJ, CxP 68532,
              Rio de Janeiro, RJ 21945-970, Brazil; e-mail: mbr@if.ufrj.br}
\begin{abstract}
This work studies the Zipf Law for cities in Brazil. Data from
censuses of 1970, 1980, 1991 and 2000 were used to select a
sample containing only cities with 30,000 inhabitants
or more. The results show that the population distribution in
Brazilian cities does follow a power law similar to the ones found in
other countries. Estimates of the power law exponent were found
to be $2.22 \pm 0.34$ for the 1970 and 1980 censuses, and $2.26 \pm
0.11$ for censuses of 1991 and 2000. More accurate results were
obtained with the maximum likelihood estimator, showing an exponent
equal to $2.41$ for 1970 and $2.36$ for the other three years.

\vspace{5.0mm}
\hspace{-3.5mm}{\it PACS:} \ 89.75Da; 89.65.Cd; 89.75.-k; 05.45.Df
\end{abstract}
\end{frontmatter}

\section{Introduction}

It was first observed by Auerbach \cite{au}, although it is often
attributed to Zipf \cite{z}, that the way in which urban aggregates are
distributed, that is, the way the populations of cities are distributed,
follows a power law behaviour with exponent $\alpha \approx 2$. If we
assign probabilities to this distribution the resulting behaviour
is also a power law, known as the {\it Zipf law}. This law seems to
have an universal character, holding at the world level \cite{zan} as
well as to single nations. The exponent also seems to be independent
of the area of the nation and the social and economical conditions
of its population \cite{mas}.

Power law exponents of cities have been measured in many countries.
It was reported by \cite{zan} that 2,400 cities in the U.S.A.\ have
$\alpha = 2.1 \pm 0.1$, whereas \cite{n05} reported $\alpha=2.30 \pm
0.05$ for the U.S.A. census of year 2000. According to \cite{zan}
1,300 municipalities in Switzerland have $\alpha=2.0 \pm 0.1$. Taking
together 2,700 cities of the world with population bigger than
100,000 inhabitants produces $\alpha=2.03 \pm 0.05$ \cite{zan}. One
should notice that those exponents were calculated by least squares
fitting, a method known to introduce biased results if data is not
properly handled \cite{g04}. Despite this, most results obtained so
far indicate that the exponent seems to follow the universal value
of $\alpha \approx 2$.

Such power law behaviour seems to be the manifestation of the
dynamics of complex systems, whose striking feature is of showing
universal laws characterized by exponents in scale invariant
distributions that happen to be basically independent of the
details in the microscopic dynamics. Social behaviour is an example
of interaction of the elements of a complex systems, in this case
human beings, giving rise to cooperative evolution which in
itself strongly differs from the individual dynamics. So, 
the demographic distribution of human beings on the Earth's
surface, which has sharp peaks of concentrated population - the
cities - alternated with relatively large extensions where the
population density is much lower, follows a power law typical of
complex system dynamics.

The aim of this paper is to present empirical evidence that the
population distribution of Brazilian cities also follows a
power law with exponent close to the universal value. We
have selected a sample from Brazil's decennial censuses of 1970,
1980, 1991 and 2000 and obtained probability distribution
functions of Brazilian cities with a lower cutoff of 30,000
inhabitants. Our procedure took great care to avoid large
statistical fluctuations at the tail in order to avoid introducing
large biases in the determination of the exponent \cite{n05,g04}.
Our results show that Brazilian cities do follow the universal
pattern: conservative estimates produced $\alpha = 2.22 \pm 0.34$
in 1970 and 1980. For 1991 and 2000 we obtained $\alpha=2.26 \pm
0.11$.

The paper is organized as follows. In \S 2 we present the data and our
selection methodology, whereas in \S 3 we present the methods to
analyze the data. \S 4 shows the results obtained using three different
techniques to calculate the exponent $\alpha$. The paper ends with a
concluding section.

\section{The Data}\lb{data}

Brazil is estimated to reach a population of approximately 185
million inhabitants by the end of 2005, the 5th place in the
ranking of the world's most populous countries. This population
occupies over 5 thousand cities, and although most of them have
very few inhabitants, 15 cities have more than one million people,
with two of them, S\~ao Paulo and Rio de Janeiro, having more than
5 million inhabitants. In order to obtain a sample for the
purposes of this work we need to define first of all what
we mean by a {\it city}. After surveying the administrative way
Brazil is governed we concluded that in Brazil's case we should
{\it equate} city to {\it municipality}, defined as being the
territorially smallest administrative subdivision of a country
that has its own democratically elected representative leadership.
This means that Brazil's entire territory is subdivided in
municipalities, or cities. Some of them have very big areas,
actually bigger than many European countries, but those are usually
located in regions very sparsely populated.

Censuses of Brazil's entire population have been taking place for over
a hundred years at a ten years hiatus since 1890. However, data in
digitalized form is only available at IBGE, the government institution
responsible for censuses, since 1970. Data in between censuses are
obtained by very small sampling and extrapolation. Considering this
we decided to take data only from the official, entire population,
censuses available in digital format, namely for the years of 1970,
1980, 1991 and 2000. This data shows that the number of Brazilian
municipalities has increased to over 30\% from 1970 to 2000. This is
clearly a consequence of the fact that the definition of a city is
administrative, reflecting Brazil's internal politics, and has been
varying over the last decades. 

The fact that the number of municipalities has shown a sharp increase
within the time span of our data will not affect our study because, as
mentioned above, most Brazilian cities have small populations and as the
Brazilian concept of a city means territorial subdivision, which includes
both rural and urban inhabitants, an examination of the data shows that
cities with more than 30 thousand inhabitants have their population
almost entirely concentrated in the urban area.\footnote{Nowadays
IBGE defines what is a rural, as opposed to an urban, area by satellite
imagery. See also footnote at page \protect\pageref{rural}.} We have,
therefore, decided to include only cities with
more than 30 thousand people in our sample, which meant a significant
reduction of the number of the municipalities as compared to the
original raw data (see table \ref{tab1}). The exclusion of the smaller
cities represents in fact the exclusion of the rural population from
our sample. In 1970 40\% of Brazilians were living in cities with less
than 30 thousand people, whereas in 2000 this figure was reduced to
26\%. In other words, roughly speaking the percentage of Brazilians
living in urban areas has increased from 60\% in 1970 to 74\% in 2000.
\begin{table}[t]
\caption{\it Number of cities in Brazil.}\lb{tab1}
\vspace{2mm}
\begin{tabular}{|c|c|c|c|c|}
\hline 
year of census & 1970 & 1980 & 1991 & 2000 \\
\hline \hline
all cities & 3958 & 3806 & 4277 & 5238 \\
\hline 
cities with $\ge$ 30,000 & 614 & 787 & 905 & 955 \\
\hline
\end{tabular}
\end{table}

\section{Data Analysis}

Once our sample is selected, we need to define our method of analysis.
Here we shall follow closely the methodology for fitting power law
distributions and estimating goodness-of-fit parameters as proposed by
\cite{n05}. We will start with a very brief introductory description
of power laws statistics in order to fix the notation.

Let ${p}(x)\: dx$ be the fraction of cities with population between $x$ and
$x+dx$. So ${p}(x)$ defines a certain distribution of the data $x$. It is
useful to express this distribution in terms of the {\it cumulative
distribution function} $\mathcal{P}(x)=\int_x^\infty
{p}(x^\prime)dx^\prime$, which is simply the probability that a city has
a population equal to or greater than $x$.
If the fraction ${p}(x)$ follows a power law of the type,
\be {p}(x) = C x^{-\alpha}, \lb{1} \ee
where $\alpha$ and $C$ are constants, then
$\mathcal{P}(x)$ also follows a power law, given by
\be \mathcal{P}(x) =
     \frac{C}{(\alpha - 1)} \; x^{-(\alpha -1)}. 
    \lb{3}
\ee
Such power law distributions are
also known as {\it Zipf law} or {\it Pareto distribution}. From
equation (\ref{1}) it is obvious that ${p}(x)$ diverges for any positive
value of the exponent $\alpha$ as $x \rightarrow 0$, and this means
that the distribution must deviate from a power law below some minimum
value $x_{\mathrm{min}}$. In other words, we can only assume that the
distribution follows a Zipf law for $x$ above $x_{\mathrm{min}}$, and
in this case equation (\ref{1}) can be normalized as $\int_{x_{\mathrm{min}
}}^\infty {p}({x^\prime}) \; d{x^\prime} =1$ to obtain the constant $C$
only if $x$ and the exponent $\alpha$ obey the following conditions:
$ \alpha > 1 $, $  x \ge x_{\mathrm{min}}$. Power laws with exponents less
than unity cannot be normalized and do not usually occur in nature \cite{n05}.
The normalized constant $C$, given in terms of $\alpha$ and $x_{\mathrm{min}}$,
allows us to write the power laws (\ref{1}) and (\ref{3}) as follows,
\be \ln p(x) = -\alpha \ln x + B, \lb{lnp} \ee
\be \ln \mathcal{P}(x) = \left( 1 - \alpha \right) \ln x + \beta,
    \lb{lnP}
\ee
where
\be B= \ln \left[ \left( \alpha -1 \right) {x_{\mathrm{min}}}^{(\alpha
       -1)} \right],
       \lb{B}
\ee
\be \beta = \left( \alpha -1 \right) \ln x_{\mathrm{min}}.
     \lb{beta}
\ee

If we now define the distribution $p(x_i)$ as being {\it the number
of cities with population equal to or bigger than $x_i$}, we are
able to create for each sample a set of $n$ observed values
$\{x_i\}, (i=1,\ldots,n), (x_1=x_{\mathrm{min}})$, from where we can
estimate $\alpha$. To do so we need first of all to create histograms
with the data once we define the step separating each set of
observed values $\{x_i\}$. The main difficulty that arises in this
procedure is the large fluctuation in the tail, towards bins which
have a far smaller number of observed values than at previous bins,
enhancing then the statistical fluctuations \cite{n05}. In order to
decrease such fluctuations we have taken logarithmic binning so that
bins span at increasingly larger intervals whose steps increase
exponentially according to the following rule, 
\be x_i= 2^{^{\scriptstyle (i-1)}}  x_{\mathrm{min}}. \lb{step} \ee
The resulting data is shown in table \ref{tab2} and plotted in figures
\ref{fig1} and \ref{fig2}, where one can clearly see a power law
behaviour for all years.\footnote{Previous attempts made by us at
plotting $\mathcal{P}(x_i)$ vs.\ $x_i$ with $x_{\mathrm{min}}<30,000$
showed no power law behaviour in Brazilian cities with population
smaller than about 25,000-30,000 inhabitants. So, the transition to
a power law behaviour does seem to indicate the change between rural
and urban population, that is, the transition from spread out human
settlements to the human population aggregations we call cities.
Hence, this cutoff in $x_i$ can be used as the critical
value that allow us to obtain the fractions of urban and rural
populations in a country.}\lb{rural} The cumulative distribution
$\mathcal{P}(x_i)$ was obtained dividing $p(x_i)$ by the total
number of cities with more than 30,000 inhabitants in each year when
an all population census occurred. This means that $\mathcal{P}(x_i)$
is the probability that a Brazilian city has population equal to or
greater than $x_i$ (see table \ref{tab1}).
\begin{table}[t]
\caption{\it Distribution functions of Brazilian municipalities.}\label{tab2}
\vspace{1mm}
\begin{tabular}{|c|c|c|c|c|c|c|c|c|c|}
 \hline
 \multicolumn{2}{|c|}{year}&\multicolumn{2}{|c|}{1970}
 &\multicolumn{2}{|c|}{1980}&\multicolumn{2}{|c|}{1991}
 &\multicolumn{2}{|c|}{2000}\\ \hline \hline
   $i$ & $x_{i}$ & $p(x_i)$ & $\mathcal{P}(x_i)$ & $p(x_i)$ &
   $\mathcal{P}(x_i)$ & $p(x_i)$ &
   $\mathcal{P}(x_i)$ & $p(x_i)$ &
   $\mathcal{P}(x_i)$  \\\hline
1 & 30,000 & 614 & 1     & 787 & 1     & 905 & 1     & 955 & 1      \\
2 & 60,000 & 187 & 0.3046& 287 & 0.3647& 383 & 0.4232& 447 & 0.4681 \\
3 & 120,000& 67  & 0.1091& 114 & 0.1449& 152 & 0.1680& 187 & 0.1958 \\
4 & 240,000& 26  & 0.0423& 45  & 0.0572& 67  & 0.0740& 92  & 0.0963 \\
5 & 480,000& 10  & 0.0163& 18  & 0.0229& 27  & 0.0298& 34  & 0.0356 \\
6 & 960,000&  5  & 0.0081& 10  & 0.0127& 12  & 0.0133& 14  & 0.0147 \\
7 & 1,920,000& 2 & 0.0033& 2   & 0.0025& 4   & 0.0044& 6   & 0.0063 \\
8 & 3,840,000& 2 & 0.0033& 2   & 0.0025& 2   & 0.0022& 2   & 0.0021 \\
9 & 7,680,000& - & -     & 1   & 0.0013& 1   & 0.0011& 1   & 0.0010 \\
\hline
   \end{tabular}
\end{table}
\begin{figure}[b]
\setlength{\unitlength}{0.240900pt}
\ifx\plotpoint\undefined\newsavebox{\plotpoint}\fi
\sbox{\plotpoint}{\rule[-0.200pt]{0.400pt}{0.400pt}}%
\begin{picture}(1650,1169)(0,0)
\font\gnuplot=cmr10 at 10pt
\gnuplot
\sbox{\plotpoint}{\rule[-0.200pt]{0.400pt}{0.400pt}}%
\put(241.0,123.0){\rule[-0.200pt]{4.818pt}{0.400pt}}
\put(221,123){\makebox(0,0)[r]{ 0.0001}}
\put(1569.0,123.0){\rule[-0.200pt]{4.818pt}{0.400pt}}
\put(241.0,188.0){\rule[-0.200pt]{2.409pt}{0.400pt}}
\put(1579.0,188.0){\rule[-0.200pt]{2.409pt}{0.400pt}}
\put(241.0,273.0){\rule[-0.200pt]{2.409pt}{0.400pt}}
\put(1579.0,273.0){\rule[-0.200pt]{2.409pt}{0.400pt}}
\put(241.0,317.0){\rule[-0.200pt]{2.409pt}{0.400pt}}
\put(1579.0,317.0){\rule[-0.200pt]{2.409pt}{0.400pt}}
\put(241.0,337.0){\rule[-0.200pt]{4.818pt}{0.400pt}}
\put(221,337){\makebox(0,0)[r]{ 0.001}}
\put(1569.0,337.0){\rule[-0.200pt]{4.818pt}{0.400pt}}
\put(241.0,402.0){\rule[-0.200pt]{2.409pt}{0.400pt}}
\put(1579.0,402.0){\rule[-0.200pt]{2.409pt}{0.400pt}}
\put(241.0,487.0){\rule[-0.200pt]{2.409pt}{0.400pt}}
\put(1579.0,487.0){\rule[-0.200pt]{2.409pt}{0.400pt}}
\put(241.0,531.0){\rule[-0.200pt]{2.409pt}{0.400pt}}
\put(1579.0,531.0){\rule[-0.200pt]{2.409pt}{0.400pt}}
\put(241.0,552.0){\rule[-0.200pt]{4.818pt}{0.400pt}}
\put(221,552){\makebox(0,0)[r]{ 0.01}}
\put(1569.0,552.0){\rule[-0.200pt]{4.818pt}{0.400pt}}
\put(241.0,616.0){\rule[-0.200pt]{2.409pt}{0.400pt}}
\put(1579.0,616.0){\rule[-0.200pt]{2.409pt}{0.400pt}}
\put(241.0,701.0){\rule[-0.200pt]{2.409pt}{0.400pt}}
\put(1579.0,701.0){\rule[-0.200pt]{2.409pt}{0.400pt}}
\put(241.0,745.0){\rule[-0.200pt]{2.409pt}{0.400pt}}
\put(1579.0,745.0){\rule[-0.200pt]{2.409pt}{0.400pt}}
\put(241.0,766.0){\rule[-0.200pt]{4.818pt}{0.400pt}}
\put(221,766){\makebox(0,0)[r]{ 0.1}}
\put(1569.0,766.0){\rule[-0.200pt]{4.818pt}{0.400pt}}
\put(241.0,830.0){\rule[-0.200pt]{2.409pt}{0.400pt}}
\put(1579.0,830.0){\rule[-0.200pt]{2.409pt}{0.400pt}}
\put(241.0,916.0){\rule[-0.200pt]{2.409pt}{0.400pt}}
\put(1579.0,916.0){\rule[-0.200pt]{2.409pt}{0.400pt}}
\put(241.0,959.0){\rule[-0.200pt]{2.409pt}{0.400pt}}
\put(1579.0,959.0){\rule[-0.200pt]{2.409pt}{0.400pt}}
\put(241.0,980.0){\rule[-0.200pt]{4.818pt}{0.400pt}}
\put(221,980){\makebox(0,0)[r]{ 1}}
\put(1569.0,980.0){\rule[-0.200pt]{4.818pt}{0.400pt}}
\put(241.0,1045.0){\rule[-0.200pt]{2.409pt}{0.400pt}}
\put(1579.0,1045.0){\rule[-0.200pt]{2.409pt}{0.400pt}}
\put(241.0,1130.0){\rule[-0.200pt]{2.409pt}{0.400pt}}
\put(1579.0,1130.0){\rule[-0.200pt]{2.409pt}{0.400pt}}
\put(241.0,123.0){\rule[-0.200pt]{0.400pt}{4.818pt}}
\put(241,82){\makebox(0,0){ 10000}}
\put(241.0,1110.0){\rule[-0.200pt]{0.400pt}{4.818pt}}
\put(364.0,123.0){\rule[-0.200pt]{0.400pt}{2.409pt}}
\put(364.0,1120.0){\rule[-0.200pt]{0.400pt}{2.409pt}}
\put(436.0,123.0){\rule[-0.200pt]{0.400pt}{2.409pt}}
\put(436.0,1120.0){\rule[-0.200pt]{0.400pt}{2.409pt}}
\put(487.0,123.0){\rule[-0.200pt]{0.400pt}{2.409pt}}
\put(487.0,1120.0){\rule[-0.200pt]{0.400pt}{2.409pt}}
\put(526.0,123.0){\rule[-0.200pt]{0.400pt}{2.409pt}}
\put(526.0,1120.0){\rule[-0.200pt]{0.400pt}{2.409pt}}
\put(559.0,123.0){\rule[-0.200pt]{0.400pt}{2.409pt}}
\put(559.0,1120.0){\rule[-0.200pt]{0.400pt}{2.409pt}}
\put(586.0,123.0){\rule[-0.200pt]{0.400pt}{2.409pt}}
\put(586.0,1120.0){\rule[-0.200pt]{0.400pt}{2.409pt}}
\put(610.0,123.0){\rule[-0.200pt]{0.400pt}{2.409pt}}
\put(610.0,1120.0){\rule[-0.200pt]{0.400pt}{2.409pt}}
\put(631.0,123.0){\rule[-0.200pt]{0.400pt}{2.409pt}}
\put(631.0,1120.0){\rule[-0.200pt]{0.400pt}{2.409pt}}
\put(649.0,123.0){\rule[-0.200pt]{0.400pt}{4.818pt}}
\put(649,82){\makebox(0,0){ 100000}}
\put(649.0,1110.0){\rule[-0.200pt]{0.400pt}{4.818pt}}
\put(772.0,123.0){\rule[-0.200pt]{0.400pt}{2.409pt}}
\put(772.0,1120.0){\rule[-0.200pt]{0.400pt}{2.409pt}}
\put(844.0,123.0){\rule[-0.200pt]{0.400pt}{2.409pt}}
\put(844.0,1120.0){\rule[-0.200pt]{0.400pt}{2.409pt}}
\put(895.0,123.0){\rule[-0.200pt]{0.400pt}{2.409pt}}
\put(895.0,1120.0){\rule[-0.200pt]{0.400pt}{2.409pt}}
\put(935.0,123.0){\rule[-0.200pt]{0.400pt}{2.409pt}}
\put(935.0,1120.0){\rule[-0.200pt]{0.400pt}{2.409pt}}
\put(967.0,123.0){\rule[-0.200pt]{0.400pt}{2.409pt}}
\put(967.0,1120.0){\rule[-0.200pt]{0.400pt}{2.409pt}}
\put(994.0,123.0){\rule[-0.200pt]{0.400pt}{2.409pt}}
\put(994.0,1120.0){\rule[-0.200pt]{0.400pt}{2.409pt}}
\put(1018.0,123.0){\rule[-0.200pt]{0.400pt}{2.409pt}}
\put(1018.0,1120.0){\rule[-0.200pt]{0.400pt}{2.409pt}}
\put(1039.0,123.0){\rule[-0.200pt]{0.400pt}{2.409pt}}
\put(1039.0,1120.0){\rule[-0.200pt]{0.400pt}{2.409pt}}
\put(1058.0,123.0){\rule[-0.200pt]{0.400pt}{4.818pt}}
\put(1058,82){\makebox(0,0){ 1e+06}}
\put(1058.0,1110.0){\rule[-0.200pt]{0.400pt}{4.818pt}}
\put(1181.0,123.0){\rule[-0.200pt]{0.400pt}{2.409pt}}
\put(1181.0,1120.0){\rule[-0.200pt]{0.400pt}{2.409pt}}
\put(1253.0,123.0){\rule[-0.200pt]{0.400pt}{2.409pt}}
\put(1253.0,1120.0){\rule[-0.200pt]{0.400pt}{2.409pt}}
\put(1304.0,123.0){\rule[-0.200pt]{0.400pt}{2.409pt}}
\put(1304.0,1120.0){\rule[-0.200pt]{0.400pt}{2.409pt}}
\put(1343.0,123.0){\rule[-0.200pt]{0.400pt}{2.409pt}}
\put(1343.0,1120.0){\rule[-0.200pt]{0.400pt}{2.409pt}}
\put(1375.0,123.0){\rule[-0.200pt]{0.400pt}{2.409pt}}
\put(1375.0,1120.0){\rule[-0.200pt]{0.400pt}{2.409pt}}
\put(1403.0,123.0){\rule[-0.200pt]{0.400pt}{2.409pt}}
\put(1403.0,1120.0){\rule[-0.200pt]{0.400pt}{2.409pt}}
\put(1426.0,123.0){\rule[-0.200pt]{0.400pt}{2.409pt}}
\put(1426.0,1120.0){\rule[-0.200pt]{0.400pt}{2.409pt}}
\put(1447.0,123.0){\rule[-0.200pt]{0.400pt}{2.409pt}}
\put(1447.0,1120.0){\rule[-0.200pt]{0.400pt}{2.409pt}}
\put(1466.0,123.0){\rule[-0.200pt]{0.400pt}{4.818pt}}
\put(1466,82){\makebox(0,0){ 1e+07}}
\put(1466.0,1110.0){\rule[-0.200pt]{0.400pt}{4.818pt}}
\put(1589.0,123.0){\rule[-0.200pt]{0.400pt}{2.409pt}}
\put(1589.0,1120.0){\rule[-0.200pt]{0.400pt}{2.409pt}}
\put(241.0,123.0){\rule[-0.200pt]{324.733pt}{0.400pt}}
\put(1589.0,123.0){\rule[-0.200pt]{0.400pt}{242.586pt}}
\put(241.0,1130.0){\rule[-0.200pt]{324.733pt}{0.400pt}}
\put(40,626){\makebox(0,0){$\mathcal{P}(x_i)$}}
\put(915,21){\makebox(0,0){$x_i$}}
\put(241.0,123.0){\rule[-0.200pt]{0.400pt}{242.586pt}}
\put(1429,1090){\makebox(0,0)[r]{1970}}
\put(436,980){\circle{18}}
\put(559,870){\circle{18}}
\put(682,774){\circle{18}}
\put(805,686){\circle{18}}
\put(928,597){\circle{18}}
\put(1050,532){\circle{18}}
\put(1173,447){\circle{18}}
\put(1296,447){\circle{18}}
\put(1499,1090){\circle{18}}
\sbox{\plotpoint}{\rule[-0.500pt]{1.000pt}{1.000pt}}%
\put(1429,1049){\makebox(0,0)[r]{1980}}
\put(436,980){\makebox(0,0){$\times$}}
\put(559,886){\makebox(0,0){$\times$}}
\put(682,800){\makebox(0,0){$\times$}}
\put(805,714){\makebox(0,0){$\times$}}
\put(928,629){\makebox(0,0){$\times$}}
\put(1050,574){\makebox(0,0){$\times$}}
\put(1173,424){\makebox(0,0){$\times$}}
\put(1296,424){\makebox(0,0){$\times$}}
\put(1419,360){\makebox(0,0){$\times$}}
\put(1499,1049){\makebox(0,0){$\times$}}
\end{picture}
\caption{\it Graph of the cumulative distribution function
$\mathcal{P}(x_i)$ against the population ${x_i}$ of Brazilian cities
with 30,000 people or more in the years of 1970 and 1980. One can
clearly see the decaying straight line pattern of a power law
behaviour with very little change over the time span of the
sample. One can also notice some fluctuations at the tail of the
plot, reflecting very small number of cities with large
population} \label{fig1}
\end{figure}
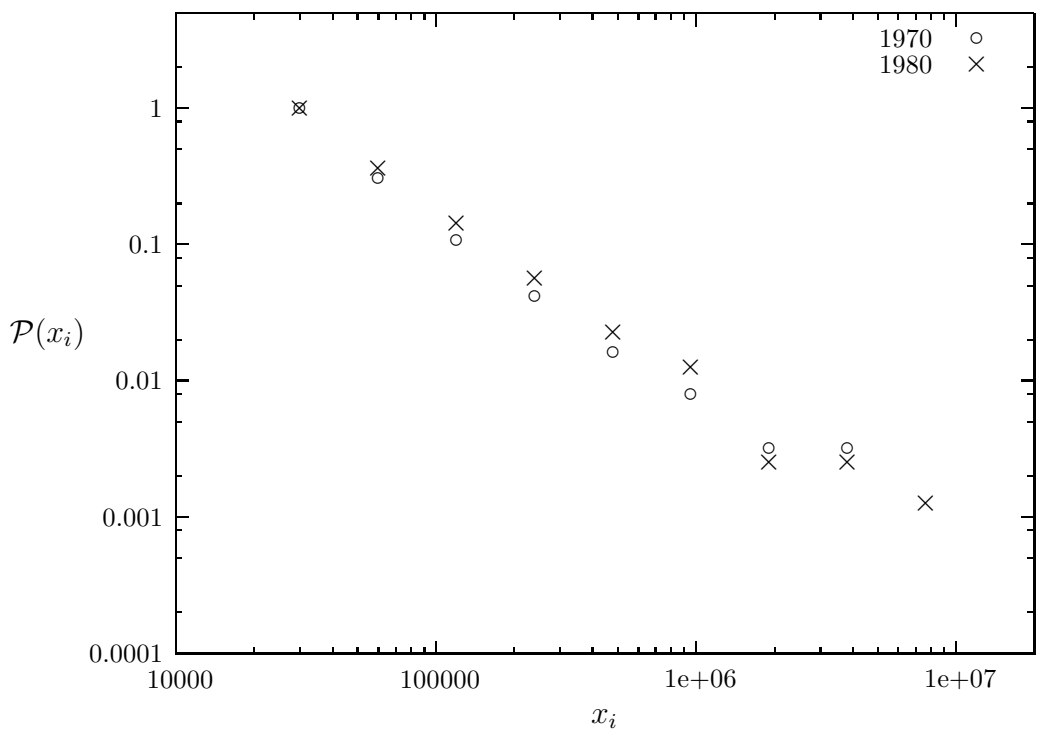
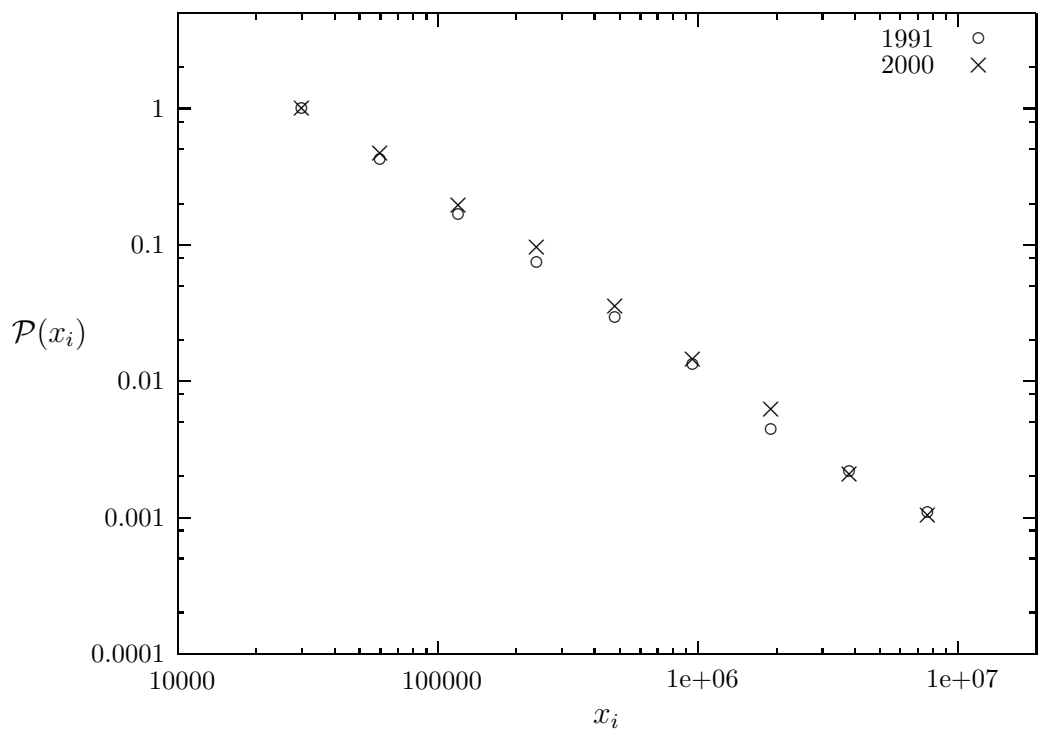
\begin{figure}[b]
\setlength{\unitlength}{0.240900pt}
\ifx\plotpoint\undefined\newsavebox{\plotpoint}\fi
\sbox{\plotpoint}{\rule[-0.200pt]{0.400pt}{0.400pt}}%
\begin{picture}(1650,1169)(0,0)
\font\gnuplot=cmr10 at 10pt
\gnuplot
\sbox{\plotpoint}{\rule[-0.200pt]{0.400pt}{0.400pt}}%
\put(241.0,123.0){\rule[-0.200pt]{4.818pt}{0.400pt}}
\put(221,123){\makebox(0,0)[r]{ 0.0001}}
\put(1569.0,123.0){\rule[-0.200pt]{4.818pt}{0.400pt}}
\put(241.0,188.0){\rule[-0.200pt]{2.409pt}{0.400pt}}
\put(1579.0,188.0){\rule[-0.200pt]{2.409pt}{0.400pt}}
\put(241.0,273.0){\rule[-0.200pt]{2.409pt}{0.400pt}}
\put(1579.0,273.0){\rule[-0.200pt]{2.409pt}{0.400pt}}
\put(241.0,317.0){\rule[-0.200pt]{2.409pt}{0.400pt}}
\put(1579.0,317.0){\rule[-0.200pt]{2.409pt}{0.400pt}}
\put(241.0,337.0){\rule[-0.200pt]{4.818pt}{0.400pt}}
\put(221,337){\makebox(0,0)[r]{ 0.001}}
\put(1569.0,337.0){\rule[-0.200pt]{4.818pt}{0.400pt}}
\put(241.0,402.0){\rule[-0.200pt]{2.409pt}{0.400pt}}
\put(1579.0,402.0){\rule[-0.200pt]{2.409pt}{0.400pt}}
\put(241.0,487.0){\rule[-0.200pt]{2.409pt}{0.400pt}}
\put(1579.0,487.0){\rule[-0.200pt]{2.409pt}{0.400pt}}
\put(241.0,531.0){\rule[-0.200pt]{2.409pt}{0.400pt}}
\put(1579.0,531.0){\rule[-0.200pt]{2.409pt}{0.400pt}}
\put(241.0,552.0){\rule[-0.200pt]{4.818pt}{0.400pt}}
\put(221,552){\makebox(0,0)[r]{ 0.01}}
\put(1569.0,552.0){\rule[-0.200pt]{4.818pt}{0.400pt}}
\put(241.0,616.0){\rule[-0.200pt]{2.409pt}{0.400pt}}
\put(1579.0,616.0){\rule[-0.200pt]{2.409pt}{0.400pt}}
\put(241.0,701.0){\rule[-0.200pt]{2.409pt}{0.400pt}}
\put(1579.0,701.0){\rule[-0.200pt]{2.409pt}{0.400pt}}
\put(241.0,745.0){\rule[-0.200pt]{2.409pt}{0.400pt}}
\put(1579.0,745.0){\rule[-0.200pt]{2.409pt}{0.400pt}}
\put(241.0,766.0){\rule[-0.200pt]{4.818pt}{0.400pt}}
\put(221,766){\makebox(0,0)[r]{ 0.1}}
\put(1569.0,766.0){\rule[-0.200pt]{4.818pt}{0.400pt}}
\put(241.0,830.0){\rule[-0.200pt]{2.409pt}{0.400pt}}
\put(1579.0,830.0){\rule[-0.200pt]{2.409pt}{0.400pt}}
\put(241.0,916.0){\rule[-0.200pt]{2.409pt}{0.400pt}}
\put(1579.0,916.0){\rule[-0.200pt]{2.409pt}{0.400pt}}
\put(241.0,959.0){\rule[-0.200pt]{2.409pt}{0.400pt}}
\put(1579.0,959.0){\rule[-0.200pt]{2.409pt}{0.400pt}}
\put(241.0,980.0){\rule[-0.200pt]{4.818pt}{0.400pt}}
\put(221,980){\makebox(0,0)[r]{ 1}}
\put(1569.0,980.0){\rule[-0.200pt]{4.818pt}{0.400pt}}
\put(241.0,1045.0){\rule[-0.200pt]{2.409pt}{0.400pt}}
\put(1579.0,1045.0){\rule[-0.200pt]{2.409pt}{0.400pt}}
\put(241.0,1130.0){\rule[-0.200pt]{2.409pt}{0.400pt}}
\put(1579.0,1130.0){\rule[-0.200pt]{2.409pt}{0.400pt}}
\put(241.0,123.0){\rule[-0.200pt]{0.400pt}{4.818pt}}
\put(241,82){\makebox(0,0){ 10000}}
\put(241.0,1110.0){\rule[-0.200pt]{0.400pt}{4.818pt}}
\put(364.0,123.0){\rule[-0.200pt]{0.400pt}{2.409pt}}
\put(364.0,1120.0){\rule[-0.200pt]{0.400pt}{2.409pt}}
\put(436.0,123.0){\rule[-0.200pt]{0.400pt}{2.409pt}}
\put(436.0,1120.0){\rule[-0.200pt]{0.400pt}{2.409pt}}
\put(487.0,123.0){\rule[-0.200pt]{0.400pt}{2.409pt}}
\put(487.0,1120.0){\rule[-0.200pt]{0.400pt}{2.409pt}}
\put(526.0,123.0){\rule[-0.200pt]{0.400pt}{2.409pt}}
\put(526.0,1120.0){\rule[-0.200pt]{0.400pt}{2.409pt}}
\put(559.0,123.0){\rule[-0.200pt]{0.400pt}{2.409pt}}
\put(559.0,1120.0){\rule[-0.200pt]{0.400pt}{2.409pt}}
\put(586.0,123.0){\rule[-0.200pt]{0.400pt}{2.409pt}}
\put(586.0,1120.0){\rule[-0.200pt]{0.400pt}{2.409pt}}
\put(610.0,123.0){\rule[-0.200pt]{0.400pt}{2.409pt}}
\put(610.0,1120.0){\rule[-0.200pt]{0.400pt}{2.409pt}}
\put(631.0,123.0){\rule[-0.200pt]{0.400pt}{2.409pt}}
\put(631.0,1120.0){\rule[-0.200pt]{0.400pt}{2.409pt}}
\put(649.0,123.0){\rule[-0.200pt]{0.400pt}{4.818pt}}
\put(649,82){\makebox(0,0){ 100000}}
\put(649.0,1110.0){\rule[-0.200pt]{0.400pt}{4.818pt}}
\put(772.0,123.0){\rule[-0.200pt]{0.400pt}{2.409pt}}
\put(772.0,1120.0){\rule[-0.200pt]{0.400pt}{2.409pt}}
\put(844.0,123.0){\rule[-0.200pt]{0.400pt}{2.409pt}}
\put(844.0,1120.0){\rule[-0.200pt]{0.400pt}{2.409pt}}
\put(895.0,123.0){\rule[-0.200pt]{0.400pt}{2.409pt}}
\put(895.0,1120.0){\rule[-0.200pt]{0.400pt}{2.409pt}}
\put(935.0,123.0){\rule[-0.200pt]{0.400pt}{2.409pt}}
\put(935.0,1120.0){\rule[-0.200pt]{0.400pt}{2.409pt}}
\put(967.0,123.0){\rule[-0.200pt]{0.400pt}{2.409pt}}
\put(967.0,1120.0){\rule[-0.200pt]{0.400pt}{2.409pt}}
\put(994.0,123.0){\rule[-0.200pt]{0.400pt}{2.409pt}}
\put(994.0,1120.0){\rule[-0.200pt]{0.400pt}{2.409pt}}
\put(1018.0,123.0){\rule[-0.200pt]{0.400pt}{2.409pt}}
\put(1018.0,1120.0){\rule[-0.200pt]{0.400pt}{2.409pt}}
\put(1039.0,123.0){\rule[-0.200pt]{0.400pt}{2.409pt}}
\put(1039.0,1120.0){\rule[-0.200pt]{0.400pt}{2.409pt}}
\put(1058.0,123.0){\rule[-0.200pt]{0.400pt}{4.818pt}}
\put(1058,82){\makebox(0,0){ 1e+06}}
\put(1058.0,1110.0){\rule[-0.200pt]{0.400pt}{4.818pt}}
\put(1181.0,123.0){\rule[-0.200pt]{0.400pt}{2.409pt}}
\put(1181.0,1120.0){\rule[-0.200pt]{0.400pt}{2.409pt}}
\put(1253.0,123.0){\rule[-0.200pt]{0.400pt}{2.409pt}}
\put(1253.0,1120.0){\rule[-0.200pt]{0.400pt}{2.409pt}}
\put(1304.0,123.0){\rule[-0.200pt]{0.400pt}{2.409pt}}
\put(1304.0,1120.0){\rule[-0.200pt]{0.400pt}{2.409pt}}
\put(1343.0,123.0){\rule[-0.200pt]{0.400pt}{2.409pt}}
\put(1343.0,1120.0){\rule[-0.200pt]{0.400pt}{2.409pt}}
\put(1375.0,123.0){\rule[-0.200pt]{0.400pt}{2.409pt}}
\put(1375.0,1120.0){\rule[-0.200pt]{0.400pt}{2.409pt}}
\put(1403.0,123.0){\rule[-0.200pt]{0.400pt}{2.409pt}}
\put(1403.0,1120.0){\rule[-0.200pt]{0.400pt}{2.409pt}}
\put(1426.0,123.0){\rule[-0.200pt]{0.400pt}{2.409pt}}
\put(1426.0,1120.0){\rule[-0.200pt]{0.400pt}{2.409pt}}
\put(1447.0,123.0){\rule[-0.200pt]{0.400pt}{2.409pt}}
\put(1447.0,1120.0){\rule[-0.200pt]{0.400pt}{2.409pt}}
\put(1466.0,123.0){\rule[-0.200pt]{0.400pt}{4.818pt}}
\put(1466,82){\makebox(0,0){ 1e+07}}
\put(1466.0,1110.0){\rule[-0.200pt]{0.400pt}{4.818pt}}
\put(1589.0,123.0){\rule[-0.200pt]{0.400pt}{2.409pt}}
\put(1589.0,1120.0){\rule[-0.200pt]{0.400pt}{2.409pt}}
\put(241.0,123.0){\rule[-0.200pt]{324.733pt}{0.400pt}}
\put(1589.0,123.0){\rule[-0.200pt]{0.400pt}{242.586pt}}
\put(241.0,1130.0){\rule[-0.200pt]{324.733pt}{0.400pt}}
\put(40,626){\makebox(0,0){$\mathcal{P}(x_i)$}}
\put(915,21){\makebox(0,0){$x_i$}}
\put(241.0,123.0){\rule[-0.200pt]{0.400pt}{242.586pt}}
\put(1429,1090){\makebox(0,0)[r]{1991}}
\put(436,980){\circle{18}}
\put(559,900){\circle{18}}
\put(682,814){\circle{18}}
\put(805,738){\circle{18}}
\put(928,653){\circle{18}}
\put(1050,578){\circle{18}}
\put(1173,476){\circle{18}}
\put(1296,411){\circle{18}}
\put(1419,346){\circle{18}}
\put(1499,1090){\circle{18}}
\sbox{\plotpoint}{\rule[-0.500pt]{1.000pt}{1.000pt}}%
\put(1429,1049){\makebox(0,0)[r]{2000}}
\put(436,980){\makebox(0,0){$\times$}}
\put(559,910){\makebox(0,0){$\times$}}
\put(682,828){\makebox(0,0){$\times$}}
\put(805,762){\makebox(0,0){$\times$}}
\put(928,670){\makebox(0,0){$\times$}}
\put(1050,587){\makebox(0,0){$\times$}}
\put(1173,508){\makebox(0,0){$\times$}}
\put(1296,406){\makebox(0,0){$\times$}}
\put(1419,341){\makebox(0,0){$\times$}}
\put(1499,1049){\makebox(0,0){$\times$}}
\end{picture}
\caption{\it Same graph as in the previous figure, but with data of 
1991 and 2000 censuses. As before, one can clearly see the decaying
straight line pattern of a power law behaviour. However, the statistical
fluctuations at the tail have virtually disappeared as compared to the
tail in figure \protect\ref{fig1}, reflecting the fact that there is a
bigger number of cities with more than one million inhabitants in Brazil
from 1991 on than in the previous years.} \label{fig2}
\end{figure}

As discussed in \S \ref{data} above, our samples assumed
$x_{\mathrm{min}}=30,000$, which still leaves $\alpha$ to be
determined. To do so we have applied three different methods to
obtain the exponent: maximum likelihood estimator, least squares
regression and parameter averaging (very simple bootstrap).
These three methods should converge to similar values
of $\alpha$, and, taken together, are capable to detect possible
systematic biases into the value of the exponent, known to
arise from simple fits from the plots (see \cite{n05,g04}). One
should notice that least squares fitting is a good method for
determining the exponent of a power law distribution,
{\it provided} the large fluctuations of the tail arising
from logarithmic binning are significantly reduced (see \cite{g04}).

\section{Results}

\subsection{Maximum Likelihood Estimator}

A simple and reliable method for extracting the exponent is to
employ the following formula discussed in \cite{n05},
\be \alpha= 1+n { \left[ \; \sum_{i=1}^n \ln \left(
            \frac{x_i}{x_{\mathrm{min}}} \right) \right] }^{-1},
    \lb{alpha}
\ee
obtained by means of the maximum likelihood estimator (MLE).
The results are shown in table \ref{tab3}, whereas figures \ref{fig3},
\ref{fig4}, \ref{fig5} and \ref{fig6} show the exponent fits of table
\ref{tab3} drawn as lines for each data.
\begin{table}[t]
\caption{\it Results for $\alpha$.}\lb{tab3}
\vspace{1mm}
\begin{tabular}{|c|c|c|c|c|}\hline
  Method & 1970 & 1980 & 1991 & 2000 \\\hline
  $\alpha_{\scriptscriptstyle \rm MLE}$ & 2,41& 2,36 & 2,36 & 2,36 \\
  $\alpha_{\scriptscriptstyle \rm LSF}$ & 2,23& 2,23 & 2,25 & 2,26 \\
  $\alpha_{\scriptscriptstyle \rm PAE}$ & 2,22 $\pm$ 0,34 &
           2,22 $\pm$ 0,34 & 2,25 $\pm$ 0,10 & 2,26 $\pm$ 0,11\\\hline
\end{tabular}
\end{table}

\subsection{Least Squares Fitting}

As noticed above, if the large uneven variation in the tail is severely
reduced, the possible bias introduced in determining the power law
exponent by least squares fitting is also reduced, as discussed in
\cite{g04}. In addition, we are applying this method together with other
two methodologies, giving us, therefore, confidence in the final
results. Results of least squares fitting (LSF) are shown in table
\ref{tab3}, whereas figures \ref{fig3}, \ref{fig4}, \ref{fig5} and
\ref{fig6} show the line fits. 

\subsection{Parameter Averaging Estimator}

\begin{figure}[b]
\input{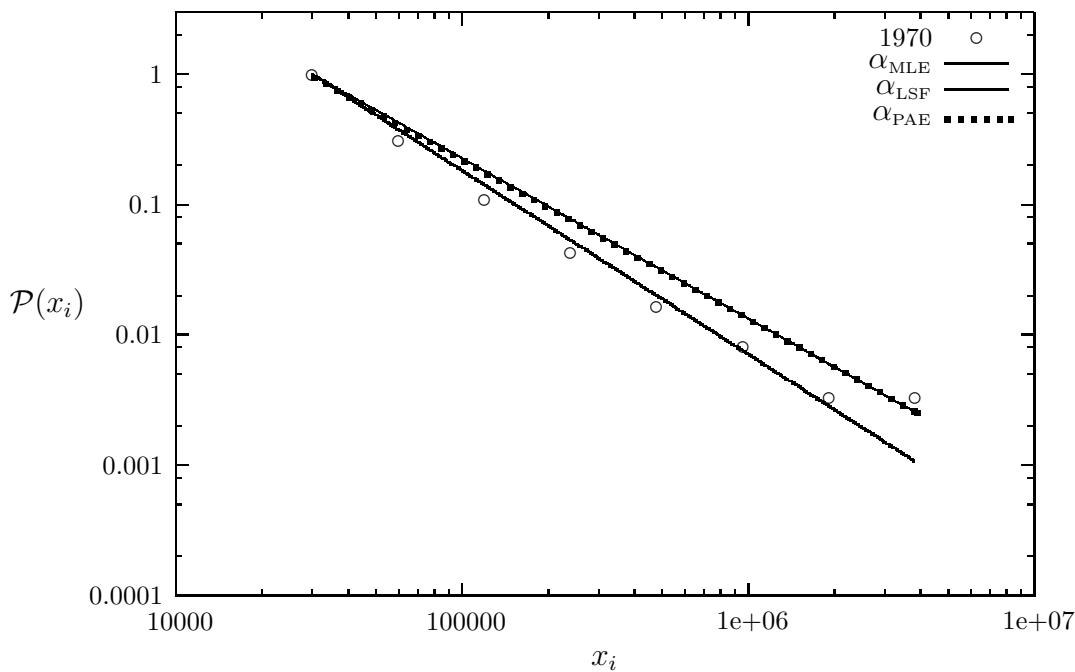}
\caption{\it Plot of $\mathcal{P}(x_i)$ vs.\ the population $x_i$ for
         1970 data with the fits shown in table \protect\ref{tab3} drawn
	 as lines. Notice that LSF and PAE estimates are almost
	 equal to one another and their line fits are superposed. In
	 addition, one can also notice that MLE does seem to provide a
	 better fit for data with larger statistical fluctuations at
	 the tail.}\label{fig3}
\end{figure}

\begin{figure}[b]
\input{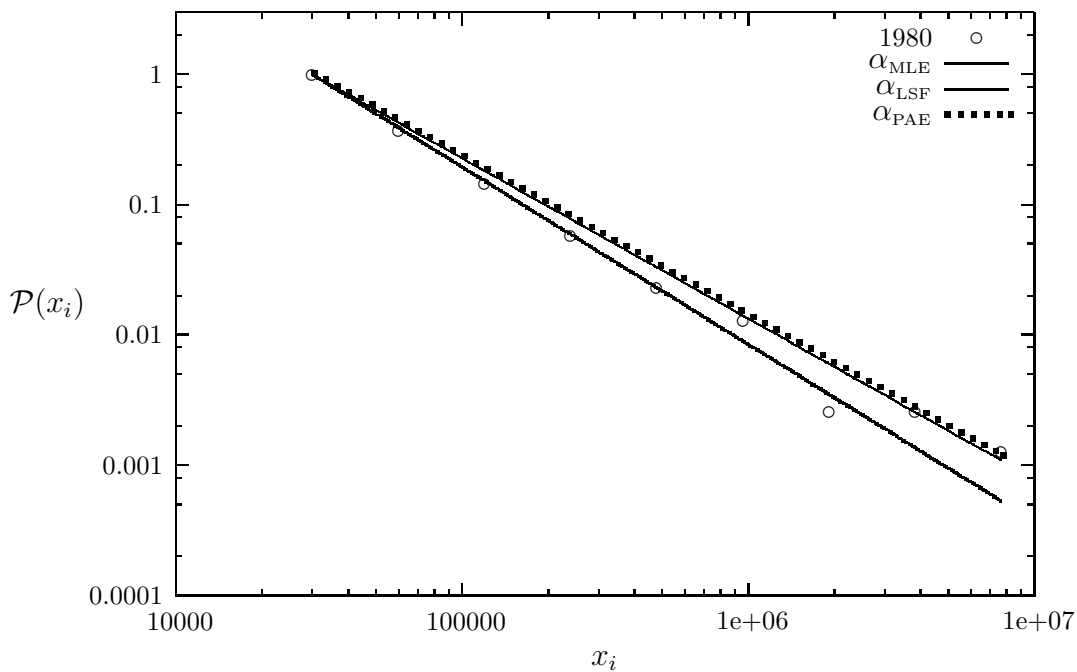}
\caption{\it Plot of $\mathcal{P}(x_i)$ vs.\ the population $x_i$ for
         1980 data with the fits shown in table \protect\ref{tab3} drawn
	 as lines. As in figure \protect\ref{fig3}, LSF and PAE results
	 are almost the same, with their line fits being drawn on top
	 of each other. Again, MLE seems to handle best the fluctuations
	 at the tail} \label{fig4}
\end{figure}

\begin{figure}[b]
\input{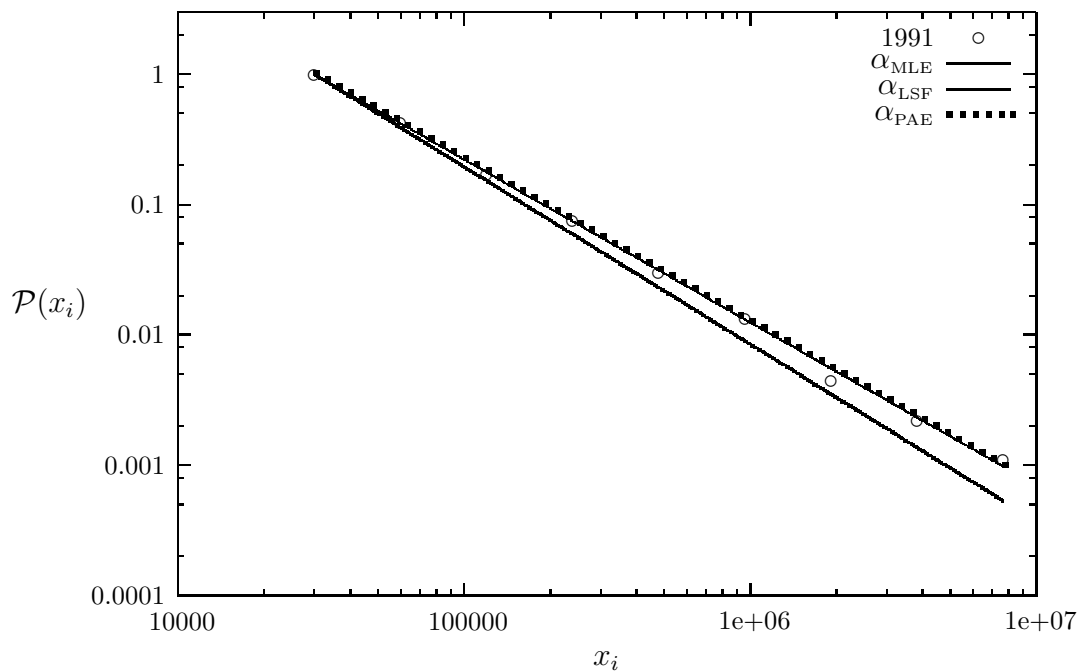}
\caption{\it Plot of $\mathcal{P}(x_i)$ vs.\ the population $x_i$ for
         1991 data with the fits shown in table \protect\ref{tab3} drawn
	 as lines. LSF and PAE results are exactly the same and the
	 exponent found with MLE is within the standard deviation of the
	 PAE result.} \label{fig5}
\end{figure}

\begin{figure}[b]
\input{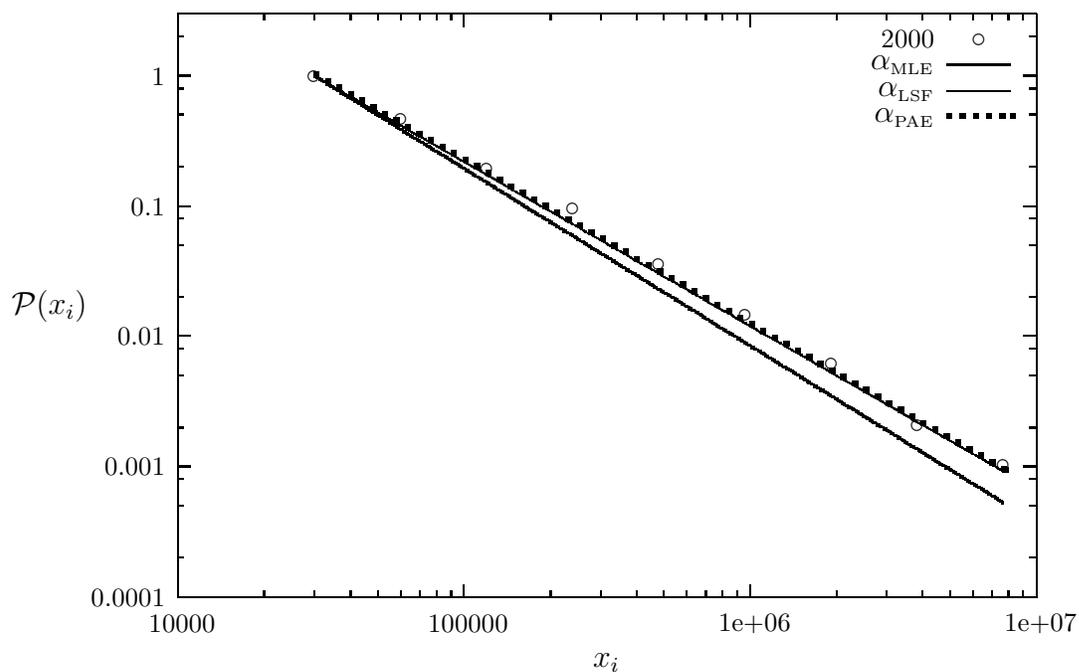}
\caption{\it Plot of $\mathcal{P}(x_i)$ vs.\ the population $x_i$ for
         2000 data with the fits shown in table \protect\ref{tab3} drawn
	 as lines. As in figure \protect\ref{fig5}, LSF and PAE results
	 are the same and MLE estimate is within PAE's standard
	 deviation. This data set is for the census with smallest
	 fluctuations at the tail as compared to the previous cases of
	 years 1970, 1980, 1991, and where all three fitting methods
	 show the smallest difference among each other (see table
	 \protect\ref{tab3}).}\label{fig6}
\end{figure}

This is in fact a very simple bootstrap estimator, where instead of
taking many random samples we have just taken all possible combinations
of two points, without repetition, obtained the angular coefficient
$\alpha$ and calculated the average and standard deviation of all values
of $\alpha$. The aim was to produce an estimate of the error.
By taking only two points we have obtained a conservative estimation
in the sense that more than two points would decrease the error.
However, viewing the results of the parameter averaging estimator (PAE)
together with the other two estimator showed us that this conservative
method is enough for the purposes of this work. The results are
also shown in table \ref{tab3} and their line fits can be found
in figures \ref{fig3}, \ref{fig4}, \ref{fig5} and \ref{fig6}. 

\subsection{Discussion}

The results obtained show that LSF and PAE estimators
produced basically the same results, whereas all MLE derived
exponents are a little higher. If we take MLE as the best estimator,
the other two suffered a bias of 8\%, 6\%, 5\% and 4\% for 1970,
1980, 1991 and 2000, respectively. Those biases are well within
the error obtained with PAE estimator, showing that once the
statistical fluctuations at the tail are successfully reduced by
means of an appropriate logarithmic binning (appropriate choice of
step and $x_{\mathrm{min}}$), LSF estimator provides a good
methodology. In fact, the bias decreases from its maximum in 1970 to
its minimum in 2000 simultaneously to a decrease in the statistical
fluctuations at the tail in these same years, brought about by the
introduction in the sample of more observed values at the tail due
to the increase in the number of cities with more than a million
inhabitants. In addition, a visual inspection of the fits in figures
\ref{fig3}, \ref{fig4}, \ref{fig5}, \ref{fig6} shows that MLE appears
to be a better fitting methodology when statistical fluctuations are
larger (1970 and 1980) as compared to smaller fluctuations in the data
stemming from the 1991 and 2000 data sets.

As an extension of our analysis it is interesting to probe why 
other authors obtain different results from the universal value
of $\alpha \approx 2$ for the power law exponent of cities, apart
from the large fluctuations at the tail and LSF fitting mentioned
above. For instance, \cite{mss} reported $\alpha \approx 1$ for
cities in Indonesia for the 1961 to 1990 decennial censuses. For
Indonesia's year 2000 census they found an exponent smaller than
one (see \cite{mss}, table 2). Inasmuch as we saw above that a
normalized power law must have $\alpha > 1$, a possible, and likely,
cause for these unexpected results is the absence of, or inappropriate, 
$x_{\mathrm{min}}$ definition for their samples. Then, without
a proper normalization it is probable that their exponent estimates
suffered contamination from the region of the plot where there is
no power law behaviour. In other words, the set of observed values
from where \cite{mss} calculated $\alpha$ was probably contaminated
with data from small cities with few inhabitants, and which should
have been removed from the data set used to calculate $\alpha$. As
seen above, finding $x_{\mathrm{min}}$ is a critical step to avoid
such a contamination. 

To summarize our results, conservative estimates for the exponent of
the Zipf law in Brazilian cities are reached by taking all methods
within the error margin. That results in $\alpha = 2.22 \pm 0.34$ for
1970 and 1980, and $\alpha=2.26 \pm 0.11$ for 1991 and 2000. On the
other hand, accurate results come from MLE estimates, producing
$\alpha=2.41$ for 1970 and $\alpha=2.36$ for the other years.

\section{Conclusion}

In this paper we have discussed the Zipf law in Brazilian cities. We
have obtained data from censuses carried out in Brazil in the years of
1970, 1980, 1991 and 2000 from where we selected a sample which included
only cities with 30,000 or more inhabitants. Then we calculated the
cumulative distribution function $\mathcal{P}(x_i)$ of Brazilian cities,
which gives the probability that a city has a population equal or bigger
than $x_i$. We found that this distribution does follow a decaying
power law, whose exponent $\alpha$ was estimated by three different
methods: maximum likelihood estimator, least squares fitting and average
parameter estimator. Our results show that a conservative estimate,
which includes the results of all three methods, produces
$\alpha = 2.22 \pm 0.34$ in 1970 and 1980, and $\alpha=2.26 \pm 0.11$
for 1991 and 2000. More accurate results are given by the maximum
likelihood estimator, showing $\alpha=2.41$ for 1970 and $\alpha=2.36$
for all other years.


\begin{thebibliography}{999}
\bibitem{au}  F. Auerbach, {\em Petermanns Geographische Mitteilungen}
        {\bf 59} (1913), 74-76
\bibitem{z}   G.K. Zipf, {\em Human Behaviour and the Principle of Least
        Effort}, Addison-Wesley, Reading, 1949
\bibitem{zan} D.H. Zanette, S.C. Manrubia, {\em Phys. Rev. Let.}
        {\bf 79} (1997) 523
\bibitem{mas} G.\ Malescio, N.V.\ Dokholyan, S.V.\ Buldyrev, H.\ Eugene
        Stanley, {\em preprint}, cond-mat/0005178 v1 (2000)
\bibitem{n05} M.E.J.\ Newman, {\em Contemporary Physics} {\bf 46} (2005)
        323, cond-mat/0412004 v2
\bibitem{g04} M.L. Goldstein, S.A. Morris, G.G. Yen,
        {\em Eur. Phys. J.} {\bf 41B} (2004) 255, cond-mat/0402322 v3
\bibitem{mss} I.\ Mulianta, H.\ Situngkir, Y.\ Surya, {\em preprint},
        nlin.PS/0409006 v1 (2004)
\end{thebibliography}
\end{document}